\documentstyle[12pt,preprint,flushrt]{aastex}

\slugcomment{\it Accepted to ApJ Letters}
\lefthead{Stern et al.}
\righthead{Near-IR Spectroscopy of SDSS~J0836+0054}

\def\eg{{e.g.,~}}
\def\etal{{et al.~}}

\def\h50min{$h_{50}^{-1}$}
\def\h65{$h_{65}^{-1}$}

\def\sdss{SDSS~J0836+0054}
\def\sdsslong{SDSS~J083643.85+005453.3}

\def\deg{\ifmmode {^{\circ}}\else {$^\circ$}\fi}

\def\secper{\ifmmode \rlap.{^{s}}\else $\rlap{.}{^{s}} $\fi}

\def\kms{\ifmmode {\rm\,km\,s^{-1}}\else
    ${\rm\,km\,s^{-1}}$\fi}
\def\kmsMpc{\ifmmode {\rm\,km\,s^{-1}\,Mpc^{-1}}\else
    ${\rm\,km\,s^{-1}\,Mpc^{-1}}$\fi}
\def\ergAcm2{\ifmmode {\rm\,ergs\,cm^{-2}\,{\rm \AA}^{-1}}\else
    ${\rm\,ergs\,cm^{-2}\,\AA^{-1}}$\fi}
\def\cm2{\ifmmode {\rm\,cm^{-2}}\else
    ${\rm\,cm^{-2}}$\fi}
\def\ergcm2s{\ifmmode {\rm\,ergs\,cm^{-2}\,s^{-1}}\else
    ${\rm\,ergs\,cm^{-2}\,s^{-1}}$\fi}
\def\cgsdeg2{\ifmmode {\rm\,ergs\,cm^{-2}\,s^{-1}\,deg^{-2}}\else
    ${\rm\,ergs\,cm^{-2}\,s^{-1}\,deg^{-2}}$\fi}
\def\ergsHz{\ifmmode {\rm\,ergs\,s^{-1}\,Hz^{-1}}\else
    ${\rm\,ergs\,s^{-1}\,Hz^{-1}}$\fi}
\def\ergs{\ifmmode {\rm\,ergs\,s^{-1}}\else
    ${\rm\,ergs\,s^{-1}}$\fi}
\def\ergsA{\ifmmode {\rm\,ergs\,s^{-1}\,\AA^{-1}}\else
    ${\rm\,ergs\,s^{-1}\,\AA^{-1}}$\fi}
\def\WHz{\ifmmode {\rm\,W\,Hz^{-1}}\else
    ${\rm\,W\,Hz^{-1}}$\fi}
\def\WHzsr{\ifmmode {\rm\,W\,Hz^{-1}\,sr^{-1}}\else
    ${\rm\,W\,Hz^{-1}\,sr^{-1}}$\fi}
\def\ergscm2Hz{\ifmmode {\rm\,ergs\,cm^{-2}\,s^{-1}\,Hz^{-1}}\else
    ${\rm\,ergs\,cm^{-2}\,s^{-1}\,Hz^{-1}}$\fi}

\def\spose#1{\hbox to 0pt{#1\hss}}
\def\simlt{\mathrel{\spose{\lower 3pt\hbox{$\mathchar"218$}}
     \raise 2.0pt\hbox{$\mathchar"13C$}}}
\def\simgt{\mathrel{\spose{\lower 3pt\hbox{$\mathchar"218$}}
     \raise 2.0pt\hbox{$\mathchar"13E$}}}

\def\lyb{Ly$\beta$}
\def\ovi{\ion{O}{6} $\lambda$1035}
\def\lya{Ly$\alpha$}
\def\nv{\ion{N}{5} $\lambda$1240}

\def\civ{\ion{C}{4} $\lambda$1549}

\def\siiii{\ion{Si}{3}] $\lambda$1892}
\def\ciii{\ion{C}{3}] $\lambda$1909}

\def\mgii{\ion{Mg}{2} $\lambda$2800}

\def\oiii{[\ion{O}{3}] $\lambda$5007}

\begin{document}

\title{Gemini-South + FLAMINGOS Demonstration Science:
\\ Near-IR Spectroscopy of the $z = 5.77$ Quasar
SDSS~J083643.85+005453.3\altaffilmark{1}}

\author{Daniel~Stern\altaffilmark{2},
Patrick~B.~Hall\altaffilmark{3,4},
L.~Felipe~Barrientos\altaffilmark{3},
Andrew~J.~Bunker\altaffilmark{5}, \\
Richard~Elston\altaffilmark{6}, 
M.~J.~Ledlow\altaffilmark{7},
S.~Nicholas~Raines\altaffilmark{6}, \&
Jon~Willis\altaffilmark{8}}

\altaffiltext{1} {Based on observations obtained at the Gemini
Observatory, which is operated by the Association of Universities for
Research in Astronomy, Inc., under a cooperative agreement with the NSF
on behalf of the Gemini partnership: the National Science Foundation
(United States), the Particle Physics and Astronomy Research Council
(United Kingdom), the National Research Council (Canada), CONICYT
(Chile), the Australian Research Council (Australia), CNPq (Brazil) and
CONICET (Argentina).}

\altaffiltext{2}{Jet Propulsion Laboratory, California Institute of
Technology, Mail Stop 169-506, Pasadena, CA 91109}

\altaffiltext{3}{Pontificia Universidad Cat\'olica de Chile, Departamento
de Astronom\'{\i}a, Facultad de F\'{\i}sica, Casilla 306, Santiago 22, Chile}

\altaffiltext{4}{Princeton University Observatory, Princeton, NJ 08544-1001}

\altaffiltext{5}{Institute of Astronomy, Cambridge, UK}

\altaffiltext{6}{Astronomy Department, University of Florida, Gainesville,
FL}

\altaffiltext{7}{Gemini-South Observatory, La Serena, Chile}

\altaffiltext{8}{European Southern Observatory, Alonso de Cordoba 3107,
Casilla 19001, Santiago 19, Chile}

\begin{abstract}

We report an infrared 1$-$1.8 $\mu$m ($J+H$-band), low-resolution ($R =
450$) spectrogram of the highest-redshift radio-loud quasar currently
known, \sdsslong, obtained during the spectroscopic commissioning run
of the FLAMINGOS multi-object, near-infrared spectrograph at the 8m
Gemini-South Observatory.  These data show broad emission from both \civ\
and \ciii, with strengths comparable to lower-redshift quasar composite
spectra.  The implication is that there is substantial enrichment of the
quasar environment, even at times less than a billion years after the
Big Bang.  The redshift derived from these features is $z = 5.774 \pm
0.003$, more accurate and slightly lower than the $z = 5.82$ reported in
the discovery paper based on the partially-absorbed \lya\ emission line.
The infrared continuum is significantly redder than lower-redshift quasar
composites.  Fitting the spectrum from 1.0~$\mu$m to 1.7~$\mu$m with a
power law $f_\nu \propto \nu^{-\alpha}$, the derived power law index is
$\alpha = 1.55$ compared to the average continuum spectral index $\langle
\alpha \rangle = 0.44$ derived from the first SDSS composite quasar.
Assuming an SMC-like extinction curve, we infer a color excess of $E(B-V)
= 0.09 \pm 0.01$ at the quasar redshift.  Only $\approx 6$\% of quasars
in the optically-selected Sloan Digital Sky Survey show comparable levels
of dust reddening.

\end{abstract}


\section{Introduction}

Quasars are amongst the most luminous objects known in the Universe,
visible at immense lookback times corresponding to when the universe
was less than a tenth its current age.  They provide ideal probes of
the conditions of the early universe, showing, for instance, that the
universe was optically thick to \lya\ at $z \simgt 6$, likely evidence
of a neutral universe seen prior to {\em an} epoch of reionization
\markcite{Becker:01, Djorgovski:01b, Kogut:03}(Becker {et~al.} 2001; Djorgovski {et~al.} 2001; Kogut {et~al.} 2003).  For the most distant
quasars, however, cosmological redshifting and the effects of
foreground hydrogen absorption make these sources largely invisible at
optical wavelengths:  the most distant quasar known currently,
SDSS~J114815.64$+$525150.3 at $z = 6.43$, shows essentially no emission
blueward of 9000 \AA\ \markcite{Fan:03}(Fan {et~al.} 2003).  Furthermore, for these most
distant systems, optical observations are largely restricted to the
rest-frame ultraviolet light around \lya.  \lya\ is a notoriously fickle
line, vulnerable to absorption from small amounts of gas and dust;
redshifts derived from \lya\ alone are systematically overestimated,
though no better feature is typically available from the discovery
optical spectra.

Near-IR observations of the highest-redshift quasars provide the
opportunity to observe complete rest-frame ultraviolet spectra of these
systems, probing out to \mgii\ for the most distant quasars.  Such
observations are essential for deriving accurate redshifts, which, in
turn, are necessary for molecular line searches \markcite{Bertoldi:03}(\eg Bertoldi \etal 2003)
and studies of the
proximity effect.  For instance, \markcite{Iwata:01}Iwata {et~al.} (2001), based on {\em ten}
days of observations with the Nobeyama Millimeter Array, report a
non-detection of CO ($J = 6 - 5$) emission in the quasar
SDSS~J104433.04$-$012502.2, assuming the redshift of $z = 5.80 \pm
0.02$ derived from the optical spectrum \markcite{Fan:00}(Fan {et~al.} 2000).  Near-IR
observations of \civ\ with the Keck telescope revised the redshift of
this quasar to $z = 5.74$ \markcite{Goodrich:01}(Goodrich {et~al.} 2001), implying that the
\markcite{Iwata:01}Iwata {et~al.} (2001) observations missed the frequency range in which CO
($J = 6 - 5$) emission would be expected.  Near-IR spectroscopy of the
highest-redshift quasars also allows estimates of the central black
hole mass in these systems \markcite{Vestergaard:02, Barth:03, Willott:03}(\eg Vestergaard 2002; Barth {et~al.} 2003; Willott, McLure, \& Jarvis 2003).
Creating black holes with masses $\simgt 10^9 M_\odot$ within a Gyr of
the Big Bang remains an outstanding astrophysical challenge
\markcite{Kauffmann:00, Wyithe:02}(\eg Kauffmann \& Haehnelt 2000; Wyithe \& Loeb 2002).

On UT 2002 November 5$-$6 and 8$-$11, we used FLAMINGOS
\markcite{Elston:98, Elston:02}(Elston 1998; Elston {et~al.} 2002) on the Gemini-South 8m telescope at Cerro
Pachon to demonstrate the infrared spectroscopic capabilities of this
unique instrument.  FLAMINGOS --- the {\bf FL}orid{\bf A} {\bf
M}ulti-object {\bf I}maging {\bf N}ear-infrared {\bf G}rism {\bf
O}bservational {\bf S}pectrometer --- is a combination wide-field
near-IR imager and multi-object spectrometer (MOS) built by the
University of Florida.  Central to the design of FLAMINGOS is a liquid
nitrogen-cooled MOS wheel, holding up to 11 MOS plates, which can be
thermally cycled in less than six hours.  FLAMINGOS is the world's
first fully-cryogenic, near-IR MOS spectrometer; it was designed to be
a peripatetic visitor instrument, traveling between observatories such
as Kitt Peak National Observatory, the Multiple Mirror Telescope, and
the southern Gemini Observatory.  At Gemini the 2048 $\times$ 2048
science-grade Hawaii-II HgCdTe array has a plate scale of 0\farcs078
pix$^{-1}$, providing a 2.5\arcmin\ $\times$ 2.5\arcmin\ field of
view.  In MOS mode it achieves spectral resolving powers of $R \sim
450$ for 0\farcs47 wide slits, covering $J + H$ bands or $H + K$ bands
simultaneously.

During this Demonstration Science run, an emphasis was placed on the
MOS mode of FLAMINGOS, though we also observed several brighter
single-slit sources to exercise the instrument.  In this {\em Letter},
we report on observations of the high-redshift quasar
\sdsslong\ \markcite{Fan:01}(hereafter \sdss; Fan {et~al.} 2001).  At the time of our
observations, \sdss, reported to be at $z = 5.82$, was the third
highest redshift quasar in the refereed literature.  With a $1.11 \pm
0.15$~mJy unresolved counterpart in the 20cm FIRST radio survey
\markcite{Becker:95}(Becker, White, \& Helfand 1995) and a Galactic absorption-corrected $0.5-2.0$~keV
flux of $F_{0.5-2.0} = 1.03 \times 10^{-14}\ \ergcm2s$
\markcite{Brandt:02}(Brandt {et~al.} 2002), \sdss\ is also amongst the most distant X-ray
sources known currently and remains the most distant radio source in
the literature.  Additional results from this Demonstration Science
observing run will be reported by Hall \etal (in preparation).


\section{Observations and Data Reductions}

We observed \sdss\ on UT 2002 November 9 using FLAMINGOS on the
Gemini-South 8m telescope.  The night was photometric and reported
0\farcs8 seeing.  We used a 0\farcs468 wide slit, the $JH$ grism, and
the $JH$ filter, providing low-resolution ($R = 450$) spectroscopy over
the wavelength range of $1.0 - 1.8~ \mu$m.  The quasar was initially
centered in the slit using $JH$ imaging.  At the time of our observing
run, Fowler sampling was not functional for the FLAMINGOS instrument,
leading to excessive readnoise.  For imaging, where the bright near-IR
sky provides background-limited data with short exposure times, this
was not a problem.  However, for spectroscopy, the effect was
deleterious in the $J$ and $H$ bands.  In order to achieve
background-limited data, we observed \sdss\ for three dithered 1500~s
integrations, unusually long exposure times for near-IR observations.
The total integration time was 4500~s and the observations were
obtained at moderate ($\sim 1.45$) airmass.

Data reductions were done with a modified version of the {\tt BOGUS}
slitmask reduction software\footnote{{\tt BOGUS-IR} is available at
{\tt http://zwolfkinder.jpl.nasa.gov/$\sim$stern/bogus.html}.} within
the {\tt IRAF} environment, and followed standard near-IR slit
spectroscopy procedures.  After multiplying by the gain of
4.1~e$^-$~DN$^{-1}$ and subtracting dark frames from the science and
calibration data, the science frames were flattened using a
median-filtered, normalized quartz lamp spectrum taken immediately
subsequent to our science observations.  Cosmic rays were then
identified and masked out.  We did a first-pass, pairwise sky
subtraction by subtracting the average of the alternate, dithered
science frames from each individual science frame.  As the near-IR sky
varied substantially over the 4500~s time frame that our data was taken
in, a second-pass sky subtraction was necessary, using a low-order fit
along the spatial axis.  The two-dimensional data, which provided good
background subtraction and background-limited data, were then combined
and the spectrum of \sdss\ was extracted.  We wavelength calibrated the
data using observations of Ar lamps obtained thru the slit immediately
subsequent to the science frames, and adjusted the wavelength zeropoint
based on telluric emission lines.

We flux calibrated the \sdss\ spectrum using spectra of the bright AOV
Hipparcos stars 113982, 112179, and 28112 obtained at low ($\sim 1.1$)
airmass.  The first of these stars was observed on the same
(photometric) night as \sdss.  The latter two stars were observed on UT
2002 November 10, a photometric night which suffered from high winds
and associated variable and poor seeing.  The resultant slit losses
were substantial and variable; we therefore normalized these data to
the UT 2002 November 9 spectrum of Hipparcos~113982, using them to
define the shape, but not the amplitude, of the sensitivity curve.  We
interpolated across the Paschen absorption features in the A0V spectra
in order to create the final sensitivity curve, which was then used to
calibrate the spectrum of \sdss, shown in Fig.~1.  We note that due to
observational limitations, all spectra were obtained at a position
angle of zero degrees, not the parallactic angle.  For optical
spectroscopy, atmospheric differential refraction will cause such a
strategy to yield erroneous measurements of spectral energy
distributions, particularly for narrow slits.  In the near-infrared,
however, this is not nearly as severe a problem.  Following
\markcite{Filippenko:82} {Filippenko} (1982), we find that atmospheric
differential refraction will compromise our observations of the
broad-band, near-infrared color of \sdss\ at only the 2$-$3\%\ level
across the $1.0 - 1.8 \mu$m wavelength region.

\section{Results}

\subsection{\ciii\ emission}

Our near-IR spectrogram provides a more accurate redshift
determination for \sdss\ than was possible from the discovery optical
spectrum, which, based on heavily-absorbed \lya\ emission and
\lyb/\ovi\ emission, provided a redshift of $z = 5.82$ \markcite{Fan:01}(Fan {et~al.} 2001).
The \ciii\ line is particularly useful for this endeavor as it is a
strong line close to the systemic redshift.  \markcite{vandenBerk:01}{Vanden~Berk} {et~al.} (2001) 
note that, unlike \lya\ and \civ, \ciii\ shows little object-to-object
variation amongst over 2200 Sloan digital sky survey (SDSS) quasar spectra they analyze.  As
a counterexample, \civ\ occasionally shows blueshifts of several
thousand \kms\ \markcite{Richards:02}(Richards {et~al.} 2002).  Fitting the \ciii\ line with a
Gaussian profile and correcting for the typical \ciii\ velocity
blueshift of 224 \kms\ relative to \oiii\ \markcite{vandenBerk:01}({Vanden~Berk} {et~al.} 2001), we calculate a redshift of
$z = 5.77$ for \sdss.  We also cross-correlated our infrared spectrum
with the SDSS composite quasar spectrum, finding a redshift of $z =
5.774 \pm 0.003$ for \sdss.  As discussed in \S 1, accurate redshifts
are essential for molecular line searches and studies of the proximity
effect.  The revised redshift implies that a substantial fraction of
the \lya\ emission line is absorbed.

Our near-IR spectrum shows a strong detection of the \ciii\ line at
1.2910 $\mu$m, with a strength of $3.0 \times 10^{-15}$ \ergcm2s.
\markcite{vandenBerk:01}{Vanden~Berk} {et~al.} (2001) present a
composite quasar spectrum derived from over 2200 Sloan digital sky
survey spectra.  In this composite, \ciii, which is slightly blended
with \siiii\ and several \ion{Fe}{3} transitions, has an equivalent
width of $21.2 $ \AA, and has a velocity width of 3700 \kms\ (FWHM).
\sdss, with a \ciii\ rest-frame equivalent width of 19 \AA\ and a
velocity width of 6300 \kms, has comparable values.  Also,
\sdss\ clearly has substantial \nv\ emission \markcite{Fan:01}(Fan
{et~al.} 2001).  These results, though not unique for high-redshift
quasars \markcite{Goodrich:01}(\eg Goodrich {et~al.} 2001), are quite
remarkable:  the emission-line gas of quasars seen when the universe
was less than a Gyr old has a similar metallicity to low-redshift
quasars.  This gas, calculated to have solar or higher metallicity
\markcite{Hamann:99, Dietrich:03}(\eg Hamann \& Ferland 1999; Dietrich
{et~al.} 2003), requires substantial {\em and} rapid enrichment.

Assuming that the broad-line region (BLR) gas dynamics is dominated by
gravitational forces, combining the radius of the BLR ($R_{\rm BLR}$)
derived from reverberation mapping studies with the velocity of the
emission line gas derived from the spectrum ($v_{\rm BLR}$), one can
derive a virial estimate of the central black hole mass, $M_\bullet =
G^{-1}~ R_{\rm BLR}~ v_{\rm BLR}^2$.  Reverberation studies find a
correlation between $R_{\rm BLR}$ and continuum luminosity
\markcite{Kaspi:00}(Kaspi {et~al.} 2000), so that black-hole mass
estimates are possible from single-epoch spectroscopic data.  However,
since reverberation mapping studies have traditionally targeted
lower-redshift, brighter sources, this relation was initially
calibrated for H$\beta$ which is inaccessible at high redshift.
Recently, \markcite{McLure:02}McLure \& Jarvis (2002) and
\markcite{Vestergaard:02}Vestergaard (2002) have calibrated the
relationship for \mgii\ and \civ, respectively, allowing black-hole
mass estimates at higher redshifts.  No \ciii\ calibration is currently
available, unfortunately, while our \civ\ detection is of insufficient
signal-to-noise to make a meaningful black-hole mass measurement.  We
note that \mgii\ resides at 1.90~$\mu$m at the redshift of \sdss, an
inaccessible wavelength for ground-based observations, residing between
the $H$ and $K$ bands.

\subsection{Red continuum}

The most striking aspect of the near-IR spectrum of \sdss\ is its red
color.  Fitting the spectrum over $\lambda \lambda\ 1.0 - 1.7~ \mu$m
(rest-frame $\lambda \lambda_0\ 1480 - 2510$~\AA) with a power law
$f_\nu \propto \nu^{-\alpha}$, we derive a power law index of 1.55,
significantly redder than the $\langle \alpha \rangle = 0.44 \pm 0.1$
continuum power-law index derived by \markcite{vandenBerk:01}{Vanden~Berk} {et~al.} (2001) for the SDSS
composite quasar for rest-frame wavelengths $\lambda \lambda_0\ 1300 -
5000$~\AA\ and redder than the $\langle \alpha \rangle = 0.57 \pm 0.33$
derived by \markcite{Pentericci:03}{Pentericci \etal} (2003) for a
sample of 45 high-redshift SDSS quasars imaged in the near-infrared.
Some of this difference derives from \ion{Fe}{2} emission
at $\lambda \lambda_0\ 2200 - 2600$~\AA\ mimicking a strong continuum.
Indeed, \markcite{Freudling:03}Freudling, Corbin, \& Korista (2003) have recently reported detection of
\ion{Fe}{2} emission in \sdss\ based on very low-resolution ($R \sim
200$) {\em Hubble Space Telescope} near-IR spectroscopy.  We note that
their spectrum is also noticeably redder than the composite quasar
spectrum it is plotted against, as well as shows \ciii\ blueshifted
relative to the composite, plotted for $z = 5.82$.  Our spectrum shows
an upturn at 1.5~$\mu$m associated with \ion{Fe}{2}.  Fitting the
\markcite{vandenBerk:01}{Vanden~Berk} {et~al.} (2001) composite over the rest-frame
wavelength range available for \sdss, we find a spectral slope of 0.8
--- significantly bluer than the spectral slope of \sdss.  We infer
that despite some contribution from \ion{Fe}{2} emission,
\sdss\ clearly has an unusually red spectrum (Fig.~1).

What is the physical significance of a red spectral slope in a
high-redshift quasar?  Reddening is the obvious implication, requiring
substantial dust in the quasar environment, at an early cosmic epoch.
The necessary processing of primordial hydrogen and helium has also
been inferred from the strong \ciii\ emission feature.  Comparing the
spectrum of \sdss\ with the SDSS composite spectrum reddened with an
SMC-like dust-reddening law \markcite{Prevot:84}(Prevot {et~al.} 1984), we estimate a color
excess of $E(B-V) = 0.09 \pm 0.01$ at the quasar redshift.  This implies an
absorption of $1.36 \pm 0.15$~mag at 1550 \AA.  \markcite{Richards:03}Richards {et~al.} (2003)
investigates the reddening of a sample of 4576 quasars from the SDSS,
finding that only $\approx$~6\% of quasars from the optically-selected
SDSS show comparable levels of dust reddening.

\markcite{Baker:95}Baker \& Hunstead (1995) report on UV/optical spectral slope variations between
various radio-loud quasar classes, placing the results in the context
of orientation-driven unification schemes.  They find that
core-dominated, radio-loud quasars have bluer spectral slopes ($\alpha
= 0.5$), albeit with very strong \ion{Fe}{2} emission.  Such quasars
are thought to be viewed along the radio-jet axis.  Lobe-dominated,
radio-loud quasars have redder optical spectra ($\alpha = 1.0$) with
weak \ion{Fe}{2} emission, suggesting large column densities of dust
obscure quasars viewed at larger inclinations to the radio-jet axis.
Compact, steep-spectrum (CSS), radio-loud quasars, however, do not fit
easily into this scheme:  their compactness would suggest they are
viewed along the radio-jet axis, but they have very red optical spectra
($\alpha = 1.4$) with no discernible 3000 \AA\ \ion{Fe}{2} bump,
suggesting substantial extinction of the central nucleus.  \sdss\ is a
radio-loud quasar with an unresolved morphology in the 20cm FIRST
survey.  Its optical spectral slope is similar to that for
the CSS composite, though the weak \ion{Fe}{2} detection suggests
perhaps a lobe-dominated system.  At present, neither multi-frequency
nor higher-resolution radio data are available for \sdss.

\section{Conclusions}

We have presented a $J + H$, low-resolution spectrum of the
high-redshift quasar \sdss, obtained during the Demonstration Science
commissioning run of the FLAMINGOS multi-object, near-IR spectrograph
at Gemini-S observatory.  We detect broad \ciii\ emission, which we use
to derive a more accurate redshift of $z = 5.774 \pm 0.003$ for this
quasar.  We also find a rest-frame ultraviolet spectral slope
significantly redder than lower-redshift quasar composites, implying
significant dust reddening of this system.  Similarly,
\markcite{Bertoldi:03} {Bertoldi \etal} (2003) have recently reported
observations of dust in the host galaxy of a quasar at $z = 6.42$.  For
the {\em Wilkinson Microwave Anisotropy Probe} cosmology
\markcite{Spergel:03}(Spergel {et~al.} 2003), the universe is slightly
less than 1~Gyr old at redshift $z = 5.77$.  Our observations of
\sdss\ require substantial enrichment of the quasar environment, even
at times only a billion years after the Big Bang.


\begin{figure}[!t]
\plottwo {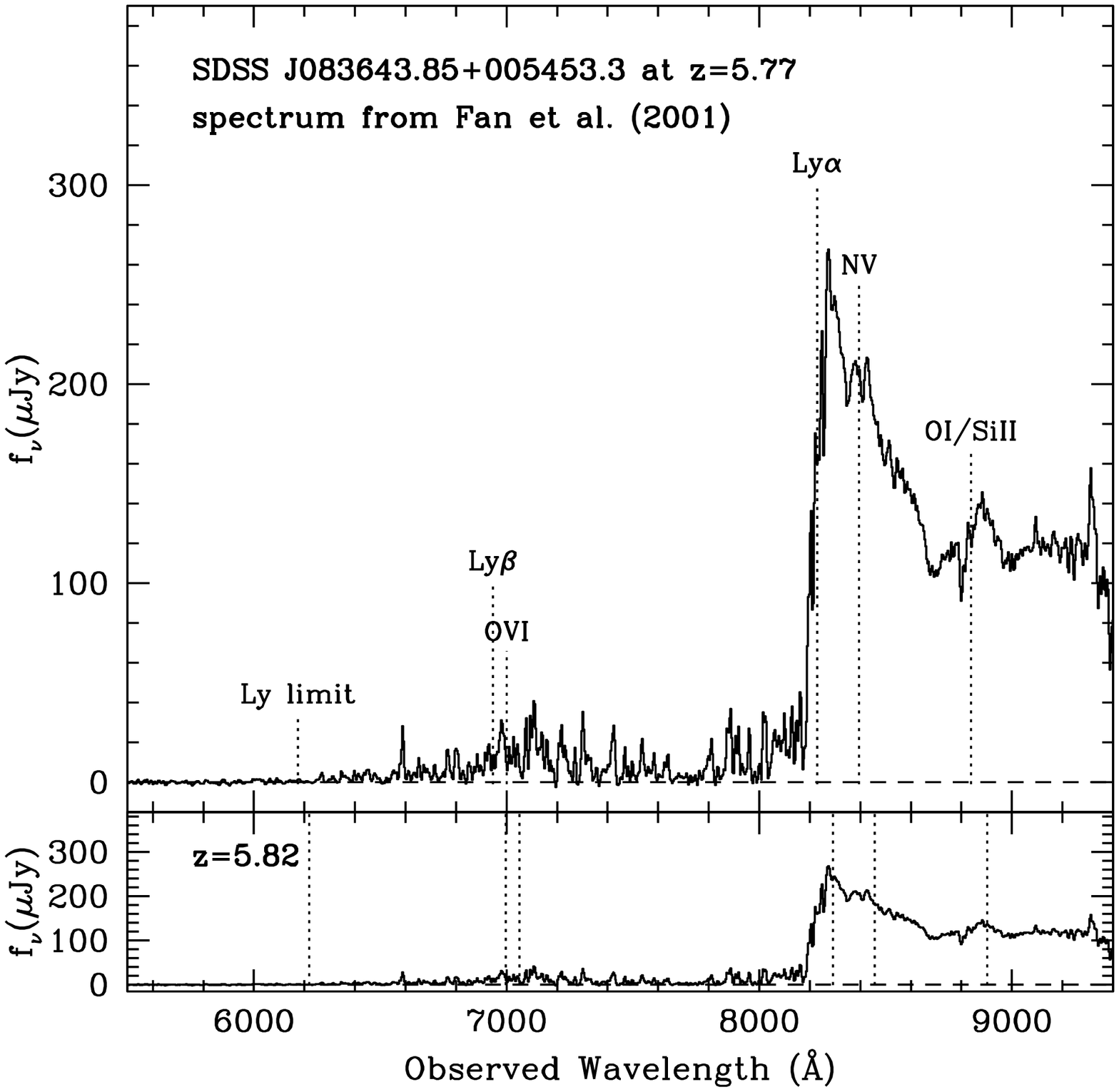}{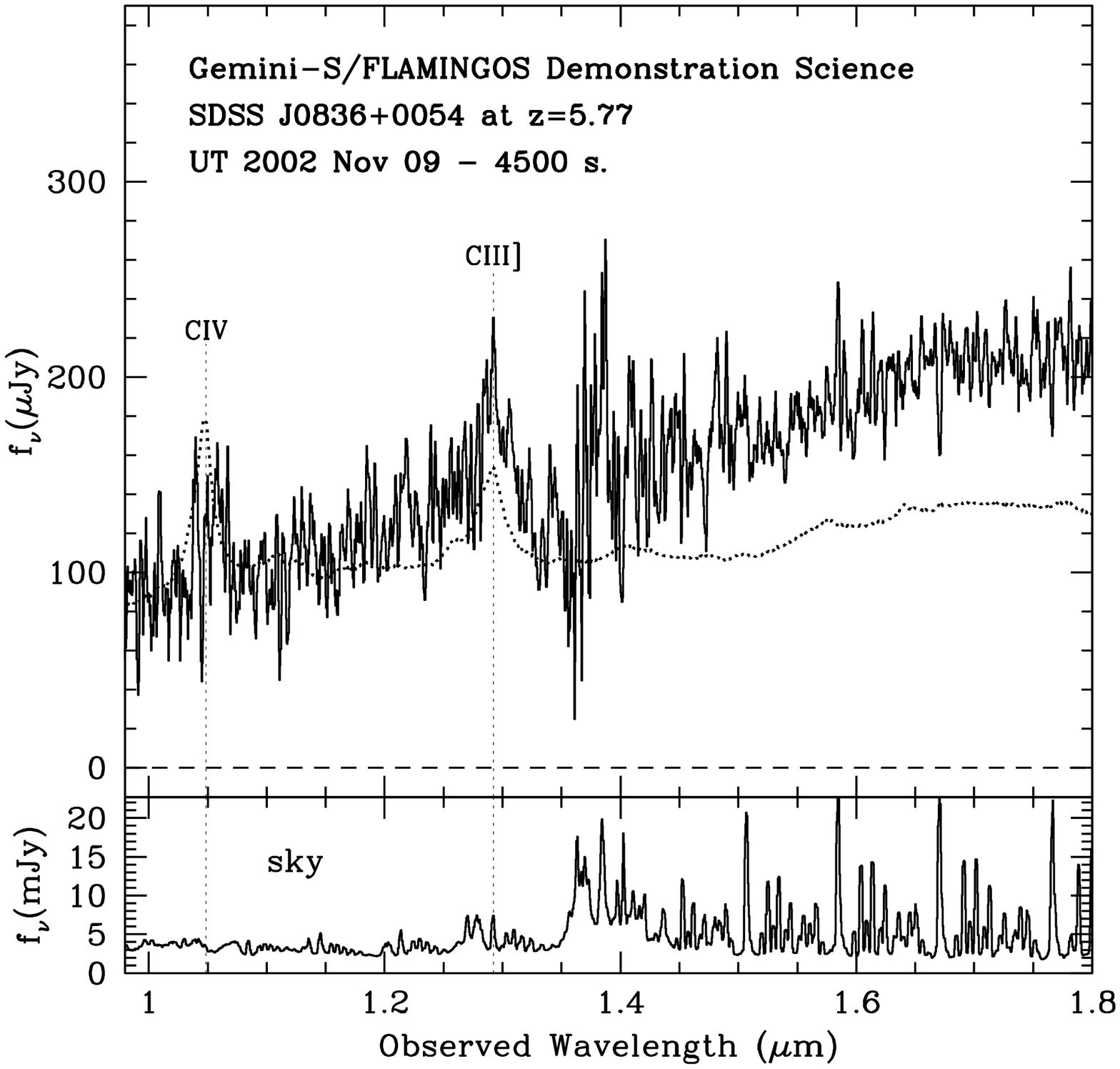}

\caption{Optical and near-IR spectra of the $z = 5.77$ quasar
SDSS~J0836+0054.  The discovery optical spectrogram, from Fan \etal
(2001) and shown in the left panels, initially yielded a redshift of $z
= 5.82$.  This redshift, shown in the lower left panel, is based on the
peak of the Ly$\alpha$ emission.  The Gemini-S/FLAMINGOS near-IR
spectrum has a good detection of the \ciii\  line, providing a more
reliable redshift for the quasar.  \civ\ is poorly detected in our
near-IR spectrogram:  most likely, this is a result of poor S/N data,
though the data also suggests \civ\ self-absorption at the quasar
redshift.  The features at 1.35$-$1.4 $\mu$m in the near-IR spectrogram
correspond to the region between the $J$ and $H$ bands, where
atmospheric transmission is poor.  The most striking aspect of the
near-IR spectrum of SDSS~J0836+0544 is its red color, discussed in \S
3.2.  For comparison, the Vanden~Berk \etal (2001) composite quasar is
shown as a dotted line.  Both spectra have been scaled to match the
published imaging photometry for SDSS~J0836+0054: $z^*_{\rm AB} = 18.74
\pm 0.05$ and $J_{\rm Vega} = 17.89 \pm 0.05$ (Fan \etal 2001).}

\end{figure}

\acknowledgements{We thank his Royal Highness, Prince Andrew, Duke of
York, for visiting the observatory on UT 2002 November 7 and inspiring
us with his regal presence.  We thank X.~Fan for the optical spectrum
of \sdss\ and A.~Barth for informative discussion.  FLAMINGOS was 
designed and constructed by the IR instrumentation group at the 
University of Florida Department of Astronomy with support from NSF
grant AST97-31180 and Kitt Peak National Observatory.  We wish to
acknowledge the many astronomers who assisted in planning the MOS
aspect to this project.  For the {\em Chandra} Deep Field-South MOS
program, we thank B.~Mobasher, A.~Koekemoer, D.~Schade, R.~Webster,
K.L.~Wu, A.~Zirm, and members of the GOODS team.   For the {\em Hubble}
Deep Field-South MOS program, we thank Ramana Athreya, David Crampton,
D.~Garcia Lambas, Leopoldo Infante, Paul Martini, Dante Minniti, Hernan
Quintana, Marcin Sawicki and Russell Smith.  The work of D.S. was
carried out at Jet Propulsion Laboratory, California Institute of
Technology, under a contract with NASA.  P.B.H. acknowledges financial
support from a Fundaci\'{o}n Andes grant.  L.F.B.'s research is
partially funded by Centro de Astrof\'{\i}sica FONDAP. }

%
%

\end{document}